\documentclass{PoS}

\usepackage{xspace}
\newcommand{\ifb}{fb$^{-1}$\xspace}
\newcommand{\pt}{$p_{\rm T}$\xspace}
\newcommand{\nb}{$N_{\rm b}$\xspace}
\newcommand{\njets}{$N_{\rm jets}$\xspace}
\newcommand{\mj}{$M_{\rm J}$\xspace}
\newcommand{\ptmiss}{$p_{\rm T}^{miss}$\xspace}
\newcommand{\htmiss}{$H_{\rm T}^{miss}$\xspace}
\newcommand{\dm}{$\Delta m$\xspace}

\newcommand{\htt}{$H_{\rm T}$\xspace}

\title{Searches for strongly-produced SUSY at CMS}

\ShortTitle{Searches for strongly-produced SUSY at CMS}

\author{\speaker{Ana Ovcharova}\thanks{on behalf of the CMS collaboration}\\
        University of California, Santa Barbara\\
        E-mail: \email{ana.ovcharova@cern.ch}}

\abstract{Searches for the pair-production of colored supersymmetric particles are presented. The results cover different scenarios of gluino and squark production, including models of split supersymmetry that predict long-lived gluinos, compressed supersymmetric mass spectra characterized by soft decay products as well as models with R-parity violation with low or no missing transverse momentum in the final state. The results are based on proton-proton collisions recorded at $\sqrt{s}=$~13~TeV with the CMS detector.}

\FullConference{ICHEP 2018, International Conference on High Energy Physics\\
        4-11 July 2018\\
        Seoul, South Korea}

\begin{document}

Since the beginning of Run 2, with the large increase in energy and then rapid increase in luminosity, the main focus of the LHC supersymmetry (SUSY) search program has been the so-called "vanilla" SUSY characterized by sizable mass splittings between the superpartners and thus signatures with significant missing transverse momentum \ptmiss. The results of these searches have been re-interpreted by the phenomenology community to assess their phase-space coverage more generally and recommend further avenues of exploration. Their findings~\cite{shih} show that even though LHC searches are approaching coverage of the full phase-space available for Natural SUSY characterized by high \ptmiss in the final states, there is significant phase-space left to explore in final states with low \ptmiss. With the LHC entering the phase of more gradual integrated luminosity scaling, the CMS experiment~\cite{cms} is continually expanding the scope of its SUSY program to probe the vast landscape of these more challenging signatures. In the following sections, we highlight three examples of how we are exploring this broader phase-space: long-lived particles in the context of Split SUSY~\cite{split1}, soft decay products in the case of SUSY models with compressed mass spectra, low \ptmiss signatures in the context of R-parity violating SUSY~\cite{mfv}.

A crucial first step towards understanding our reach and limitations in exploring the long-lived particle frontier is to assess the sensitivity of our prompt searches to various displaced particle decays. For this purpose we re-interpret a search for prompt gluino pair production~\cite{alphat} which aims to cover a maximal phase-space characterized by three main requirements: at least 1 jet, no leptons and transverse hadronic missing energy \htmiss greater than 200~GeV. To increase sensitivity to various models, the phase-space is binned in hadronic transverse energy \htt, \htmiss, jet multiplicity \njets and b-jet multiplicity \nb, resulting in 254 search regions. The multijet background in this search is reduced to less than 1\% with variables designed to reject events with fake \ptmiss due to instrumental effects. Remaining electroweak backgrounds are estimated by extrapolating yields from dedicated control regions in data. 

\begin{figure}
\centering
    \includegraphics[width=0.6\textwidth]{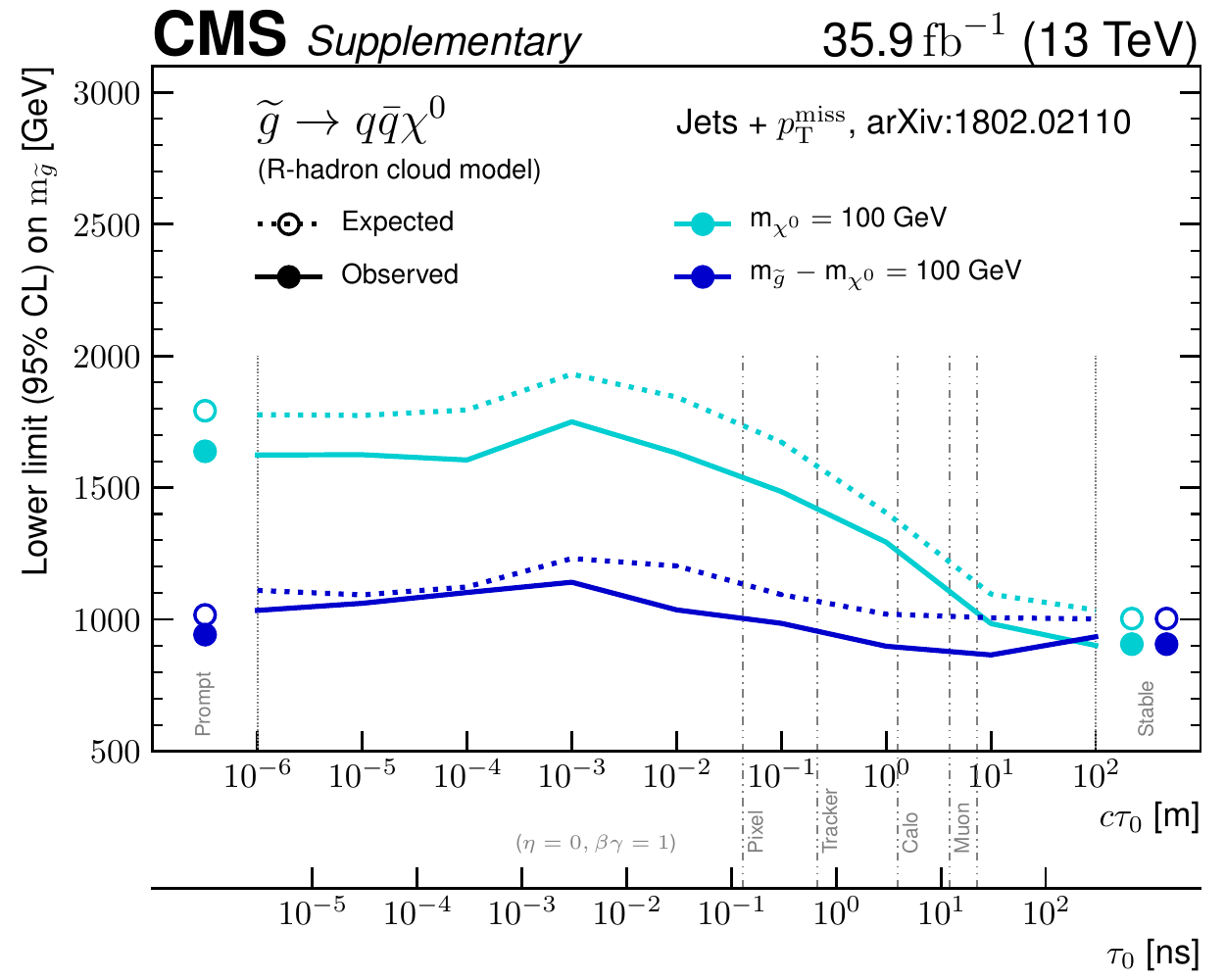}
 \caption{Re-interpretation of a search for pair-production of gluinos with prompt decays in the context of Split SUSY with displaced gluino decays~\cite{alphat}.}
 \label{fig:alphat}
 \end{figure}

Figure~\ref{fig:alphat} shows the results on the re-interpretation of this search in the context of Split SUSY~\cite{split1}, where signatures with displaced jets naturally arise from the decays of long-lived gluinos. As expected from the high \ptmiss search requirements, the sensitivity for  gluinos with short lifetimes is marked by significantly higher reach for high mass splittings (\dm) between the gluino and the lightest supersymmetric particle (LSP). The bump in sensitivity at lifetimes of approximately 1 mm is due to displaced vertices being identified by the b-quark tagging algorithm. At even longer lifetimes, the sensitivity in the large \dm scenario gradually degrades due to inefficiency of the jet cleaning cuts used to reject instrumental backgrounds when applied to the displaced jet. For displacements of 10 m and beyond, the jet with highest transverse momentum (\pt) is no longer a result of the gluino decay but rather from prompt  initial-state radiation (ISR) and thus the sensitivity flattens. In the case of small \dm, due to the high \ptmiss requirements, the leading jet is generally due to ISR and thus the sensitivity is flat as a function of the lifetime. Thanks to various features of the final state we find that even conventional searches for promptly decaying gluinos have sensitivity to the entire lifetime spectrum all the way to stable gluinos.
Further steps in expanding our coverage of signatures with long-lived particles require specialized searches exploiting the striking features associated with these final states such as disappearing tracks, displaced vertices and many more. The latest results from such searches are covered in a separate dedicated contribution. 

Another way to weaken limits from conventional SUSY searches is to have a compressed mass spectrum where no significant energy is available to boost the LSP, resulting in little or no \ptmiss in the final state. To illustrate one approach in tackling this difficult final state, we highlight a search for stop pair production~\cite{soft1l}, targeting mass splittings between the stop and the LSP as low as 10~GeV. The final state is characterized by a single lepton with low transverse momentum \pt and low \ptmiss. Firstly, to allow for triggering on the signal events, we require the presence of a jet with $p_T> 100$ GeV in order to boost the stop pair system and thus enhance the \ptmiss in the events. Then, to further separate signal from backgrounds, we take advantage of the markedly soft \pt spectrum of the lepton in the signal as compared to the background. To maximize the signal efficiency, the lepton \pt thresholds are pushed down to 3.5~GeV for muons and 5~GeV for electrons. An additional discrimination can be gained also from exploiting the transverse mass of the lepton and the \ptmiss which tends to have a shape that is distinct from the shape of the W boson transverse mass distribution seen in background processes. We attempt to maximally exploit the data by exploring two approaches, one commonly referred to as cut-and-count (CC) and the other based on multivariate analysis (MVA). In the CC approach the events are  binned in the powerful variables mentioned above, while the MVA is trained on the detailed event kinematics of the signal as a function of \dm in 10 GeV increments, looking at a number of variables involving the lepton and jets kinematics. Since the MVA is optimized for a particular signal model, it has better reach for the targeted 4-body stop decay for certain \dm scenarios, while the CC approach is more versatile for the purposes of re-interpretation in additional models such as a chargino-mediated stop decay. 

Prompt lepton background estimates are based on simulation normalized to the data in control regions at low BDT score and high lepton \pt for the MVA and CC approach, respectively. Non-prompt lepton  backgrounds, important for phase space with high lepton-\ptmiss transverse mass and very low lepton \pt, are extrapolated from control regions with looser lepton isolation via an efficiency for fakes derived in data. The search excludes stops below approximately 400 to 560 GeV depending on the stop-LSP mass splitting, which is varied from 10 to 80 GeV. In contrast, searches at high \dm where the final state is characterized by high \ptmiss, and therefore more distinct signature from the backgrounds, exclude stops up to about 1100~GeV~\cite{stop_himet}.

Finally, another set of models that provide solution to the hierarchy problem, yet evade the limits of conventional high-\ptmiss searches, are SUSY spectra involving R-parity violation. The following search~\cite{rpvmj} concentrates on the scenario referred to as Minimal Flavor Violation~\cite{mfv} where the stop quark can decay to two quarks via a B-number violating vertex. Since the couplings in this model are associated with the standard model Yukawa couplings, the preferred decay is to bottom and strange quarks. Specifically, we consider gluino pair production with the gluino decaying to a combination of a top, a bottom and a strange quark. The search strategy involves taking advantage of the large hadronic activity for trigger and background discrimination. We require hadronic transverse energy exceeding 1200 GeV and at least 6 jets. To further exploit the event structure marked by correlations between the jets originating from the decay of a heavy parent particle, we recluster events into large-radius jets and use the scalar sum of all jet masses \mj to discriminate between the background and the signal. To improve sensitivity to various signal masses the phase-space is binned in \njets and \mj. The signal is then extracted via a global fit of of the \nb distribution in bins of \njets and \mj. The \nb shape for each background is taken from simulation. The nuisances constraining the \nb shape are based on studies of b-tagging efficiency and gluon-splitting modeling in data, as well as constraints on the background normalization from dedicated control regions for each process in data in bins of \njets and \mj. The data is found to be in good agreement with the background-only hypothesis and the search excludes gluinos up to 1610~GeV. Similarly to the stop search above, we can compare this mass reach to the corresponding high-\ptmiss signature, where we exclude gluino masses as high as $\sim$1950~GeV~\cite{gluino_himet}.

In summary, we have excluded gluinos and squarks up to about 1100~GeV and 1950~GeV, respectively, in various final states with high \ptmiss, thus posing a significant challenge to Vanilla SUSY and the CMS SUSY program is actively expanding to cover available phase-space characterized by more exotic as well as more challenging final states. The examples given here represent a small set of analyses from the rich search program performed with the first 36~\ifb of LHC Run 2 data. The significantly larger luminosity accumulated by the end of 2018 will allow us to push our reach in the realm of high-\ptmiss final states as well as to further expand our discovery potential along these new avenues.

\end{document}